\documentclass[%
 reprint,
 amsmath,amssymb,
 aps,
]{revtex4-2}

\usepackage{xcolor}
\usepackage{hyperref}
\hypersetup{
  colorlinks,
  citecolor=blue,
  linkcolor=blue,
  urlcolor=black}

\usepackage{booktabs}
\usepackage{graphicx}
\usepackage{dcolumn}
\usepackage{float}
\usepackage{bm}

\begin{document}

\preprint{APS/123-QED}

\title{Linear Magnetoresistance and Anomalous Hall Effect in the Superconductor NiBi$_{3}$}

\author{Gabriel Sant'ana}
\email{gabriel.santana.dasilva@outlook.com}
\author{Jully Paola Peña Pacheco}
\author{David Mockli}
\author{Fabiano Mesquita da Rosa}
\author{Sergio Garcia Magalhães}
\author{Paulo Pureur}
\author{Milton A. Tumelero}
 \email{matumelero@if.ufrgs.br}
\affiliation{ 
Instituto de F\'{i}sica, Universidade Federal do Rio Grande do Sul, 91501-970 Porto Alegre, Brazil}

\date{\today}

\begin{abstract}
The NiBi$_{3}$ compound exhibits a compelling interplay between superconductivity and magnetism, further enriched by topological characteristics that make it an exceptional platform for exploring emergent electronic phenomena. Here, we report experimental evidence of unconventional magnetic phenomena in high-quality single crystals of NiBi$_3$, revealed through detailed magnetotransport measurements. The magnetoresistance displays a non-usual temperature dependence, featuring both a classical Lorentz-like component and a linear-in-field contribution. In addition, anomalous Hall effect signal persist down to the superconducting transition temperature and vanish above 75 K. These observations suggest that magnetic fluctuations play a significant role in charge transport in NiBi$_{3}$, highlighting a magnetically and topologically intertwined electronic structure. Our findings underscore the complex and multifaceted nature of this material.

\end{abstract}

\maketitle

The interplay between superconductivity, magnetism, and topology has driven the discovery of several emergent electronic phases. Some of these states offer potential for future energy and information technologies through the development of magnetic and topological superconductors, as well as spintronic devices \cite{linder2015superconducting}. 
The intermetallic compound NiBi$_3$ is a type-II superconductor with strong electron-phonon coupling and a critical temperature of approximately 4 K \cite{kumar2011physical,silva2013superconductivity}. Early studies suggested coexistence of ferromagnetism (FM) and superconductivity due to magnetization observed below the critical temperature \cite{pineiro2011possible,herrmannsdorfer2011structure}. Nevertheless, first principle calculations have indicated a non-magnetic ground state and investigations on the magnetization of single crystals have pointed out that high temperature long-range magnetic ordering in this material is rather related to Nickel inclusions \cite{shang2023fully,wang2023absence}. Evidences of low temperature ordering \cite{herrmannsdorfer2011structure} and results on electron spin resonance signals, detected below 150 K \cite{zhu2012surface}, keep open the prospect of unconventional magnetic order in this compound, which may be related to the presence of magnetic fluctuations \cite{zhu2012surface}, low-dimensional ordering and even compensated magnetization. Still, the presence of a magnetic phase in the NiBi$_{3}$ as well as its implication in the superconducting state remains under open debate. 

Recent studies combining density functional theory and angle-resolved photoemission spectroscopy (ARPES) reveals topological surface states in NiBi$_3$ single crystal \cite{adriano2023bulk,zhang2025}. Although topological superconductivity was ruled out—since the surface bands lie far below the Fermi energy, additional features such as nodal lines and nodal surfaces make the physics of this system even more rich.

Additionally, epitaxial Ni/Bi bilayers have demonstrated signatures of unconvention superconductivity, with presence of chiral $p$-wave symmetry \cite{gong2015possible,gong2017time,chauhan2019nodeless,wang2017anomalous,hayashi2024,cai2023,he2022}. In these systems, the spontaneous formation of NiBi$_3$ at the interface appears to play a central role in the onset of superconductivity \cite{vaughan2020origin,siva2015spontaneous,sant2024ni,Park2024,chao2019superconductivity}. Indeed, the role of NiBi$_3$ has been identified as critical to the development of unconventional pairing mechanisms \cite{liu2018superconductivity,Park2024}. However, several microscopic studies on pure NiBi$_3$ - including muon spin rotation \cite{shang2023fully}, Andreev reflection spectroscopy \cite{zhao2018singlet}, and Kerr rotation experiments \cite{wang2023absence} have consistently ruled out the possibility of triplet pairing \cite{gati2018effect}.

Despite the recent progress in the field, the normal-state magnetic and electronic properties of NiBi$_3$ remain largely unexplored. Most importantly, the correspondence between the electronic properties of NiBi$_{3}$ and the Bi/Ni bilayer remains unclear. Here, we investigate the magnetotransport behavior of NiBi$_3$ by using high-quality single crystals. Our results reveal an unconventional magnetoresistance that is linear in magnetic field at low fields and increases linearly with temperature. Moreover, we report, for the first time, Hall effect measurements for this compound, which display a substantial anomalous contribution. Together, our findings point to strong spin fluctuations in the electronic properties in NiBi$_3$, which could be fundamental to describe the superconducting properties of NiBi$_3$ and Bi/Ni systems.



\begin{figure}[h!]
    \centering
    \includegraphics[width=8.5cm]{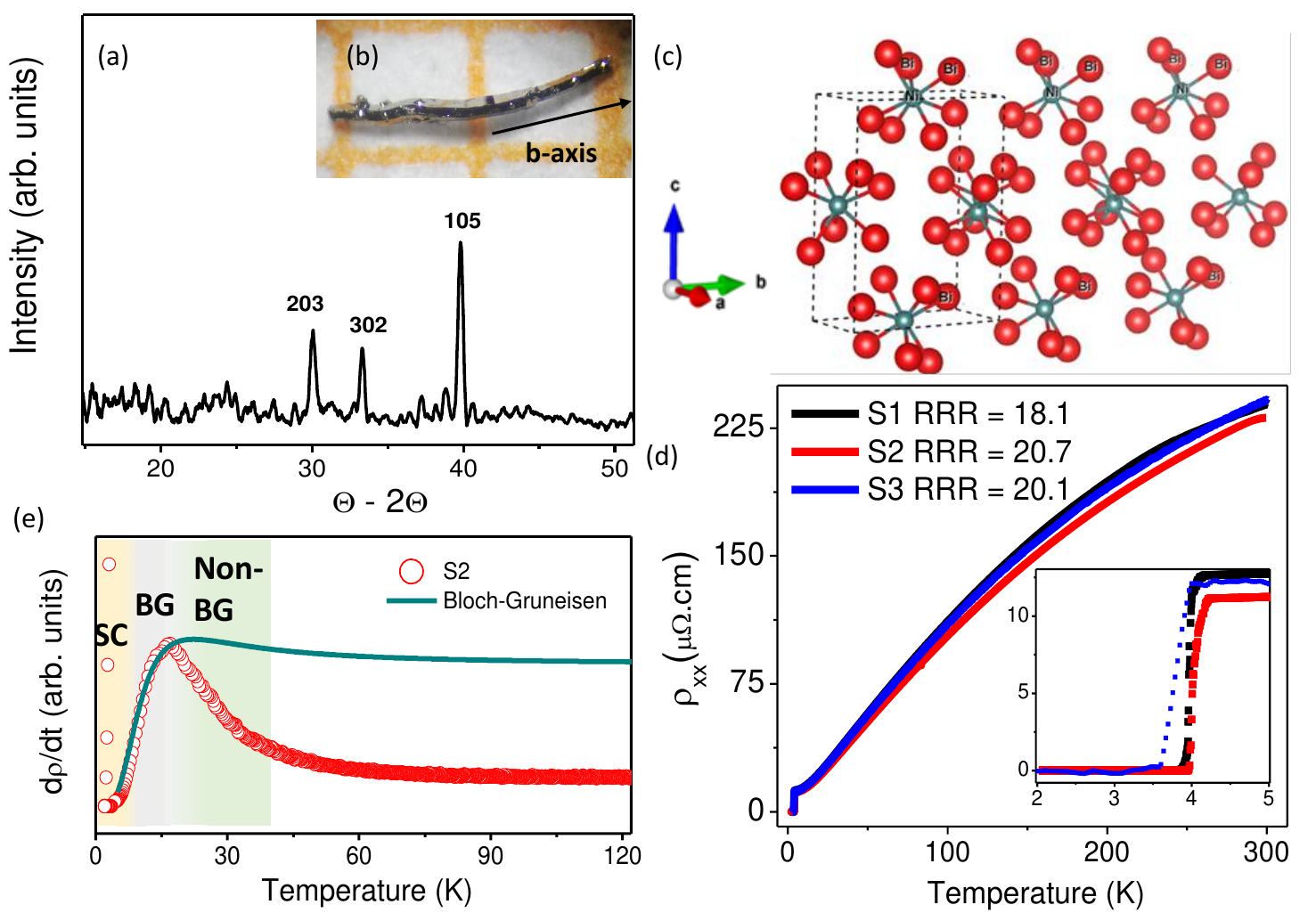}
    \caption{(a) GIXRD diffractogram from several NiBi$_3$ needle-like single crystals. (b) Photograph of a single crystal. (c) Side view of the orthorhombic NiBi$_3$ structure. (d) Temperature-dependent resistivity for samples S1, S2, and S3. \textit{Inset:} Zoom-in of the superconducting transition. (e) First derivative of resistivity with respect to temperature for sample S2. Solid lines represent simulations using the Bloch–Grüneisen (BG) model.}
    \label{fig:GIXRD_Resistance}
\end{figure}

Needle-like NiBi$_3$ samples were prepared by the self-flux method using a component ratio of Bi$_{9}$:Ni$_1$. High-purity Bi (99.999$\%$) and Ni (99.99$\%$) powders were sealed in a vacuum quartz ampule, heated to 1150$^\circ$C, and subjected to a two-step cooling process: initially cooled from 1100$^\circ$C to 740$^\circ$C at a rate of 3$^\circ$C/min, followed by slow cooling to 400$^\circ$C at 1.6$^\circ$C/h. Single crystals were separated from the flux by centrifugation. 
Figure~\ref{fig:GIXRD_Resistance}{\color{blue}b} shows a representative needle-like crystal with approximate dimensions of 2 mm$\times$ 0.1 mm $\times$ 0.1 mm. Grazing incidence X-ray diffraction (GIXRD) measurements were carried out using Cu$\alpha_1$ radiation to determine phase purity and crystallographic orientation.
Magnetotransport measurements were performed using a home-built setup in a CMag9 cryostat (Cryomagnetics) and in a 9T PPMS system (Quantum Design), employing the standard four-probe method. The electrical current was applied along the crystallographic $b$-axis.

Fig. \ref{fig:GIXRD_Resistance}{\color{blue}a} shows the GIXRD pattern obtained from a batch of NiBi$_3$ single crystals. The diffraction peaks correspond to the orthorhombic NiBi$_3$ phase with $Pnma$ symmetry, as indicated in the figure. No secondary or segregated phases were detected in the XRD analysis. Furthermore, all crystals were aligned along the $b$-axis, as evidenced by the exclusive observation of reflections with $k = 0$ in the Miller indices, confirming preferential needle-like growth along the crystallographic $b$-direction.

The atomic structure of NiBi$_3$ is schematically illustrated in Fig. \ref{fig:GIXRD_Resistance}{\color{blue}c}. NiBi$_3$ consists of long quasi-one-dimensional chains extending along the $b$-axis. The unit cell contains two such chains, each comprising two Ni atoms. These chains are not equivalent: one is rotated relative to the other by approximately 25° about the $b$-axis. Despite the structural quasi-1D character, the valence bands exhibit significant energy dispersion \cite{zhang2025}.

Fig. \ref{fig:GIXRD_Resistance}{\color{blue}d} displays the temperature dependence of the longitudinal resistivity ($\rho_{xx}$) for three different NiBi$3$ single crystals. The observed profiles are consistent with previous reports for this material, featuring early resistivity saturation, a superconducting transition near 4 K (see inset of Fig. \ref{fig:GIXRD_Resistance}{\color{blue}d}), and residual resistivity ratios (RRRs) in agreement with literature values \cite{shang2023fully,silva2013superconductivity,zhu2012surface,fujimori2000superconducting,wang2023absence}. The residual resistivities ($\rho_{0}$) were measured as 12.7, 11.2, and 12.1 $\mu\Omega\cdot$cm for samples S1, S2, and S3, respectively, yielding RRRs (defined as $\rho$(300 K)/$\rho_{0}$) of 18.1, 20.7, and 20.1. All samples exhibit sharp superconducting transitions with widths below 300 mK.

\nocite{*}

Fig. \ref{fig:GIXRD_Resistance}{\color{blue}e} shows the resistivity derivative for sample S2; results for other samples are provided in the Supplementary Material (SM). Between 5 K and 15 K, the resistivity follows the standard Bloch–Grüneisen (BG) behavior, described by:

\begin{equation}
    \rho_{BG}(T) = \rho_{0} + A\left(\frac{T}{\Theta_{D}}\right)^{5}\int_{0}^{\frac{\Theta_{D}}{T}}\frac{x^{5}\mathrm{d}x}{(e^{x} - 1)(1 - e^{-x})},
\end{equation}
where $\Theta_D$ is the Debye temperature and $A$ is a material-dependent constant. The BG model accounts for carrier scattering by acoustic phonons. The simulated curve is shown alongside the experimental data in Fig. \ref{fig:GIXRD_Resistance}{\color{blue}e} (gray-shaded region). A fit yields $\Theta_D \approx 65$ K, consistent with previous experimental estimates \cite{fujimori2000superconducting, zhu2012surface}.

According to the BG model, the first derivative of resistivity initially increases as $T^4$, reaching a maximum that marks the crossover where phonon population becomes significant. Beyond this point, the derivative slightly decreases and approaches a plateau, marking the regime where electron-phonon scattering saturates. In our experimental data, the maximum in $d\rho/dT$ occurs near 16 K for sample S2. Above this temperature, however, the derivative decays exponentially, deviating from BG predictions (green-shaded region in Fig. \ref{fig:GIXRD_Resistance}{\color{blue}e}). This deviation strongly suggests an additional scattering mechanism beyond electron-phonon coupling.

To date, the resistivity behavior in NiBi$_3$ has been modeled using various approaches, including parallel resistor models \cite{fujimori2000superconducting,nedellec1985anomalous}, exponential terms \cite{zhu2012surface,sakurai2000thermoelectric}, and Bloch–Grüneisen functions \cite{shang2023fully} to account for the early resistivity saturation. However, a microscopic description of the transport mechanism remains unavailable.

\begin{figure}[h]
    \centering
    \includegraphics[width=8.5cm]{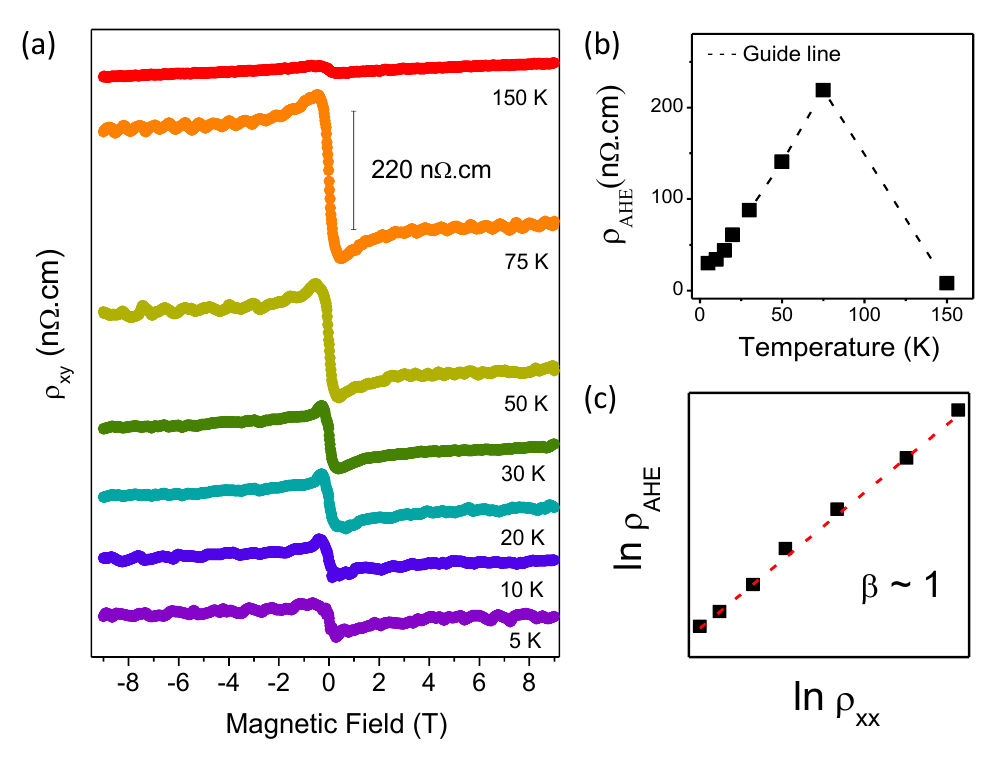}
    \caption{Anomalous Hall effect in NiBi$_3$ (sample S3). (a) $\rho_{xy}$ versus $H$ at various temperatures. (b) $\rho_{\mathrm{AHE}}$ as a function of temperature; the gap between 80 and 140 K highlights the abrupt drop around 150 K. (c) ln plot of $\rho_{\mathrm{AHE}}$ vs $\rho_{xx}$ reveling the linear relation among the quantities.}
    \label{fig:anomalous}
\end{figure}

Fig. \ref{fig:anomalous}{\color{blue}a} shows transverse resistivity measurements ($\rho_{xy}$) in NiBi$_3$, clearly exhibiting anomalous Hall effect (AHE) features. The S-shaped $\rho_{xy}$ versus $H$ curves are characteristic of AHE: while $\rho_{xy}$ saturates at high fields (ordinary Hall regime), it reverses abruptly near zero field. The anomalous Hall resistivity, $\rho_{\mathrm{AHE}}$, is extracted by extrapolating the linear high-field region to zero field. At low temperatures, $\rho_{\mathrm{AHE}}$ is approximately 25 n$\Omega\cdot$cm, increasing to about 220 n$\Omega\cdot$cm near 75 K. Above 150 K and up to 300 K, the AHE diminishes significantly, nearly vanishing.

The temperature dependence of $\rho_{\mathrm{AHE}}$ is summarized in Fig. \ref{fig:anomalous}{\color{blue}b}, while Fig. \ref{fig:anomalous}{\color{blue}c} shows the natural logarithmic plot of $\rho_{\mathrm{AHE}}$ versus $\rho_{xx}$, indicating a linear scaling consistent with $\rho_{\mathrm{AHE}} \propto \rho_{xx}^{\beta}$.

The AHE is typically associated with broken time-reversal symmetry (TRS) \cite{nagaosa2010anomalous}, most commonly arising from ferromagnetism (FM) due to uncompensated magnetic moments. However, the presence of long-range static FM order in NiBi$_3$ has been extensively ruled out \cite{zhu2012surface,silva2013superconductivity}. Indeed, our SQUID magnetometry measurements confirm the absence of spontaneous magnetization (see SM, Fig. {\color{blue}SM2}, showing FC and ZFC curves under various magnetic fields). Notably, AHE has also been observed in compensated spin systems such as altermagnets \cite{PhysRevX.12.040501,PhysRevLett.133.086503,PhysRevB.110.094425} and non-collinear antiferromagnets \cite{smejkal2022}, where $\rho_{\mathrm{AHE}}$ emerges despite a net-zero magnetic moment. However, first-principles calculations for NiBi$_3$ indicate a non-magnetic ground state, which may exclude antiferromagnetic (AFM) coupling \cite{zhang2025, kumar2011physical}. Still, the AHE has also been reported in non-magnetic materials exhibiting strong spin fluctuations \cite{spinfluct1}, such as the intermetallic compound SmMnBi$_2$ \cite{spinfluct2}.

Furthermore, the observed linear scaling of $\rho_{\mathrm{AHE}}$ with $\rho_{xx}$ (with $\beta \sim 1$) is typical of magnetic systems and is commonly attributed to skew-scattering mechanisms \cite{smit1955spontaneous, nagaosa2010anomalous}. A subtle feature in $\rho_{xy}$ near zero field may also indicate the presence of additional scattering effects.

\begin{figure}[h]
    \centering
    \includegraphics[width=8cm]{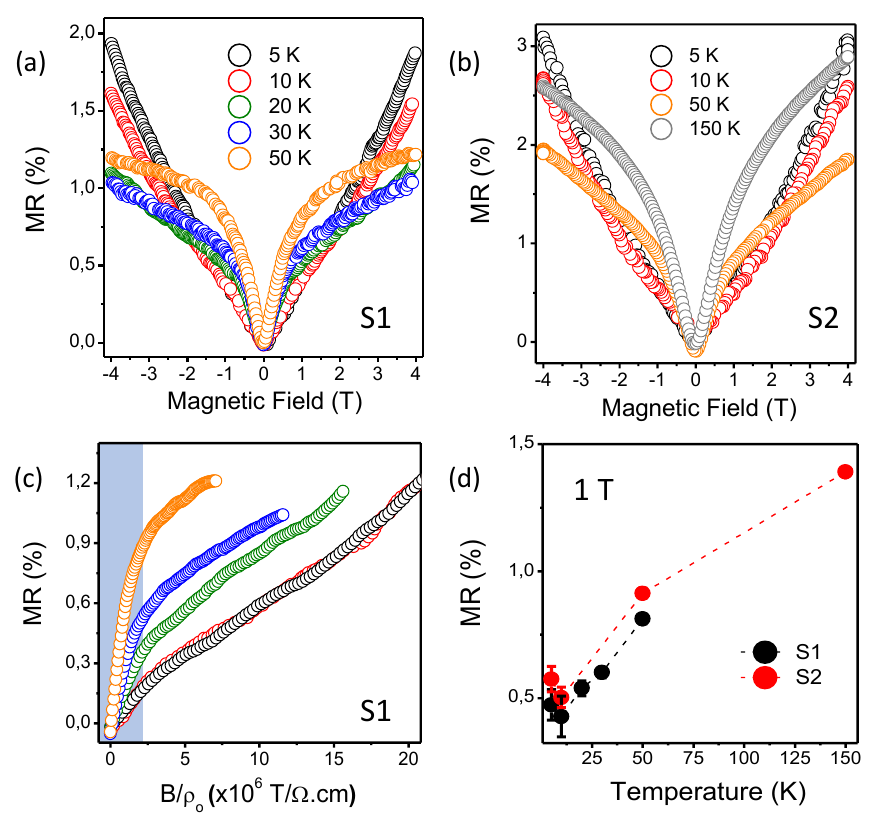}
    \caption{Magnetoresistance (MR) in function of magnetic field for (a) S1 and (b) S2. (c) MR in function of temperature at 1 T for all samples. (d) Kohler's rule analysis for samples S1.}
    \label{fig:MR_1der_kohler}
\end{figure}

Fig. \ref{fig:MR_1der_kohler}{\color{blue}a,b} shows the longitudinal magnetoresistance ($\mathrm{MR}$), defined as $\Delta\rho(B)/\rho_0$, measured in the normal state with a and parallel magnetic field applied along the $b$-axis. Data for additional samples are provided in the SM (Fig. {\color{blue}SM3}). At 5 K and 10 K, the $\mathrm{MR}$ displays a parabolic-like behavior with positive curvature, consistent with orbital deflection of charge carriers. Above 20 K, the MR remains positive but exhibits negative curvature and becomes linear at low fields. The $\mathrm{MR}$ magnitude initially decreases with temperature, reaching a minimum between 10 K and 20 K, and then increases again at higher temperatures. Fig. \ref{fig:MR_1der_kohler}{\color{blue}d} summarizes the $\mathrm{MR}$ values for various samples and temperatures.

The Kohler plot for sample S1 is shown in Fig. \ref{fig:MR_1der_kohler}{\color{blue}c}. At low temperatures, the curves collapse, obeying Kohler’s rule (KR). However, at elevated temperatures, clear deviations from KR are observed. The semiclassical KR relation assumes a single conduction channel with isotropic, field-independent scattering rates \cite{kohler1938magnetischen}:

\begin{equation}
    \mathrm{MR} = f(\omega_{c}\tau) = f\left(\frac{B}{\rho(0,T)}\right),
\end{equation}
where $\omega_c$ is the cyclotron frequency, $\tau$ is the scattering time, and $\rho(0,T)$ is the zero-field resistivity. Deviations from KR in NiBi$_3$ have been previously reported \cite{zhu2012surface}. In general, such violations are attributed to multiband transport or anisotropic scattering mechanisms. Although a temperature-dependent carrier density ($n_c$) has been proposed as a possible cause \cite{xu2021extended}, our data suggest otherwise. In particular, the linear increase in $\mathrm{MR}$ at low fields above 20 K (blue-shaded region in Fig. \ref{fig:MR_1der_kohler}{\color{blue}c}) prevents any scaling collapse with the lower-temperature curves. This indicates that changes in $n_c$ are not the primary origin of KR violation in NiBi$_3$. Notably, the temperature at which the MR behavior changes coincides with the maximum in $d\rho/dT$ (see Fig. \ref{fig:GIXRD_Resistance}{\color{blue}e}), reinforcing the presence of a secondary scattering mechanism. As discussed below, the rapid decay of the resistivity derivative, the emergence of the AHE, and the change in low-temperature MR behavior are all likely related to spin fluctuations in this system.

\begin{figure}
    \centering
    \includegraphics[width=8.5cm]{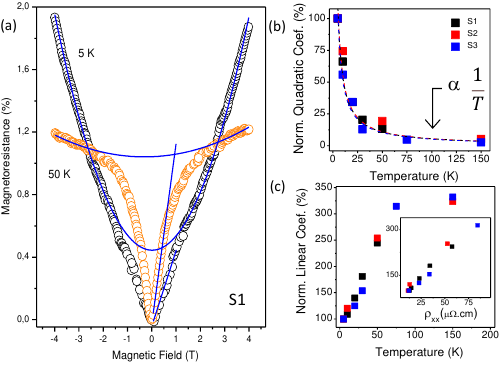}
    \caption{(a) Fitting of MR into low-field linear and high-field quadratic components. (b) Quadratic coefficient normalized at 5 K vs. temperature for S1, S2, and S3. (c) Temperature-dependence of the linear coefficient normalized at 5 K. \textit{Inset}: its correlation with $\rho_{xx}$.}
    \label{fig:fitting_analsysis}
\end{figure}

To further investigate the magnetotransport mechanism, we fitted the MR curves in two regimes: a quadratic component at high fields and a linear term at low fields. Fig. \ref{fig:fitting_analsysis}{\color{blue}a} illustrates the fitting procedure for sample S1 at 5 K and 50 K. The extracted quadratic coefficients, normalized to their 5 K values, are plotted in Fig. \ref{fig:fitting_analsysis}{\color{blue}b} as a function of temperature for samples S1, S2, and S3. These coefficients reflect the ordinary MR, which scales as $\mathrm{MR} \propto (\mu B)^2$, where $\mu$ is the carrier mobility, typically inversely proportional to the resistivity in metals. The dashed lines in Fig. \ref{fig:fitting_analsysis}{\color{blue}b} represent a $1/T$ trend, which agrees well with the data, indicating that ordinary MR dominates at low temperatures and high magnetic fields.

The low-field linear MR coefficient (for $B < 1$ T) is shown in Fig. \ref{fig:fitting_analsysis}{\color{blue}c}, exhibiting a monotonic increase with temperature for all samples. Notably, this linear term scales proportionally with the longitudinal resistivity, in stark contrast to the quadratic contribution, as revealed in the inset of Fig. \ref{fig:fitting_analsysis}{\color{blue}c}. While several mechanisms have been proposed to explain high-field linear magnetoresistance (LMR), only a few studies have addressed the low-field regime, such as in density wave systems \cite{pnas.1820092116}. However, none of these frameworks satisfactorily accounts for the behavior observed in NiBi$_3$.

The simultaneous presence of linear LMR and the AHE provides compelling evidence for magnetic phenomena in NiBi$_3$. To interpret the LMR in terms of spin fluctuations, we employed a modified two-current model originally proposed by Campbell and Fert \cite{A_Fert_1976}. In this framework, spin-flip scattering mixes the two spin channels, introducing an additional resistive component.
In our modified formulation, the magnetoresistance is described by:
\begin{equation}
    \frac{ \Delta \rho(H) }{ \rho(0) } = C \cdot \frac{H}{H_0 + H}, \label{eqn:MR_fert}
\end{equation}
where $C$ represents the high-field magnetoresistance limit, and $H_0$ is the characteristic saturation field. A full derivation of this expression and details of the model are provided in the SM. The core assumption is that spin-flip processes are enhanced by the magnetic field within the spin fluctuation regime.

Eq. \ref{eqn:MR_fert} accurately reproduces the experimental data in the low-field regime and captures the observed linear behavior. In NiBi$_3$, the spin fluctuation regime extends below 150 K, as revealed by both the AHE data and the temperature dependence of the carrier density (see SM, Fig. {\color{blue}SM2}). Notably, both the AHE and LMR are dynamic phenomena governed by ultrafast scattering processes, with characteristic timescales ranging from $10^{-15}$ to $10^{-12}$ s. These timescales indicate that such effects can probe spin fluctuations that persist longer than the electronic scattering time. Finally, our results demonstrate that both LMR and AHE persist down to the superconducting transition temperature, revisiting the question of the conventional nature of superconductivity in NiBi$_{3}$ and also in Bi/Ni bilayers, now from the standpoint of its underlying spin fluctuations.



In conclusion, the observation of the anomalous Hall effect and low-field linear magnetoresistance below 150 K provides strong evidence of magnetism in NiBi$_3$. These findings complement previous reports of magnetic fluctuations and topological features, contributing to the understanding of the complex electronic structure of this material. In particular, the linear magnetoresistance whose amplitude scales with the longitudinal resistivity appears to be intrinsically linked to an underlying magnetic order, highlighting an unconventional transport mechanism that warrants further investigation.

\begin{acknowledgments}
The authors would like to acknowledge the funding agencies FAPERGS, CAPES and CNPq, in terms of the following grants: Pronex grant no. 16/0490-0 (Fapergs-CNPq)
\end{acknowledgments}

\bibliography{apssamp}

\section*{Supplementary Materials}

\renewcommand{\thefigure}{SM\arabic{figure}}
\setcounter{figure}{0}

\begin{figure}[h]
    \centering
    \includegraphics[width=7cm]{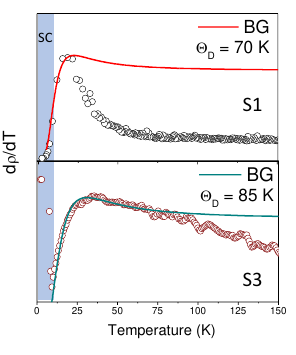}
    \caption{First derivative of resistivity for S1 and S3. Colored lines correspond to the simulation of Bloch-Guruneisen (BG) first derivative for a fixed Debye temperature ($\Theta_{D}$). Blue-shaded region divergence is due to the superconducting (SC) transition.}
    \label{fig:BG_samples}
\end{figure}

The first derivative of the resistivity with respect to temperature, d$\rho$/dT, is shown in Fig. \ref{fig:BG_samples} for samples S1 and S3. The sharp divergence within the blue-shaded region corresponds to the superconducting transition. Above the critical temperature, both samples exhibit behavior consistent with the Bloch-Grüneisen (BG) model, as indicated by the solid colored lines. The temperature at which phonon modes become thermally activated—marked by the peak in d$\rho$/dT indicates the crossover to dominant electron-phonon scattering. For SQ, this peak occurs around 15 K, similar to the behavior of sample S2 discussed in the main text, while for S3 the peak appears near 30K. Beyond the peak, the decrease in d$\rho$/dT reflects a deviation from the linear temperature dependence predicted by the BG model. This deviation occurs smoothly for S3 but more abruptly for S1, again resembling the behavior observed in S2. The contrast is attributed to variations in crystal size and morphology, which affect the Debye temperature ($\Theta_D$), estimated as 70 K for S1 and around 85 K for S3.

\begin{figure}
    \centering
    \includegraphics[width=1\linewidth]{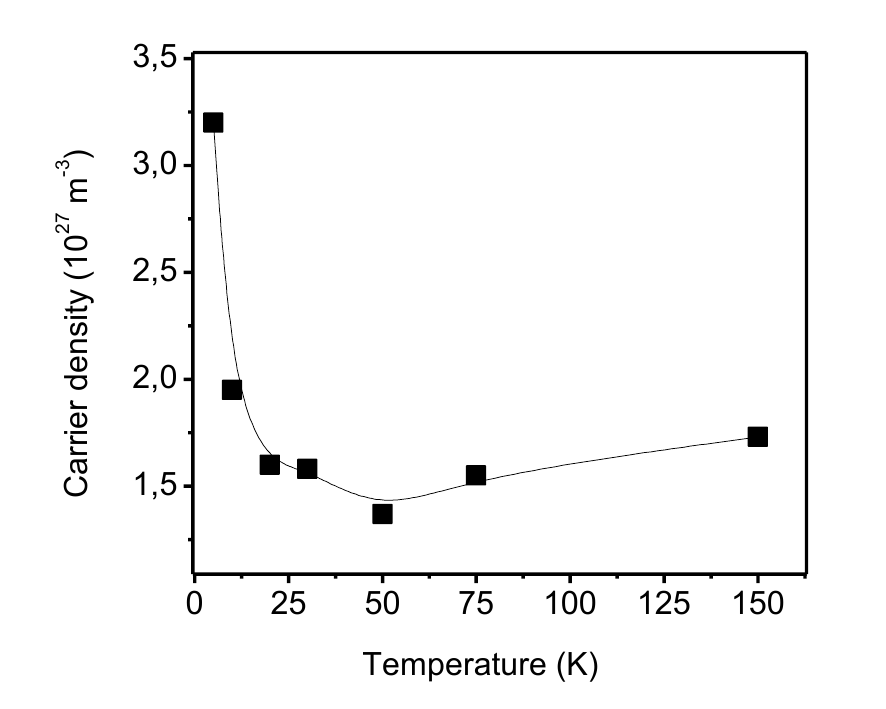}
    \caption{Carrier density in function of temperature.}
    \label{fig:carrier_density}
\end{figure}

The carrier density for the NiBi$_{3}$ is presented in Fig. \ref{fig:carrier_density}. The carrier density was obtained by linear fitting the R$_{xy}$ vs. B curves at high field, where the AHE are saturated. The curve displays a fast decay of carrier density, reaching to a minima at about 50 K. This low temperature carrier suppression can be related to spin  fluctuations in the system.

\begin{figure}[h]
    \centering
    \includegraphics[width=8.5cm]{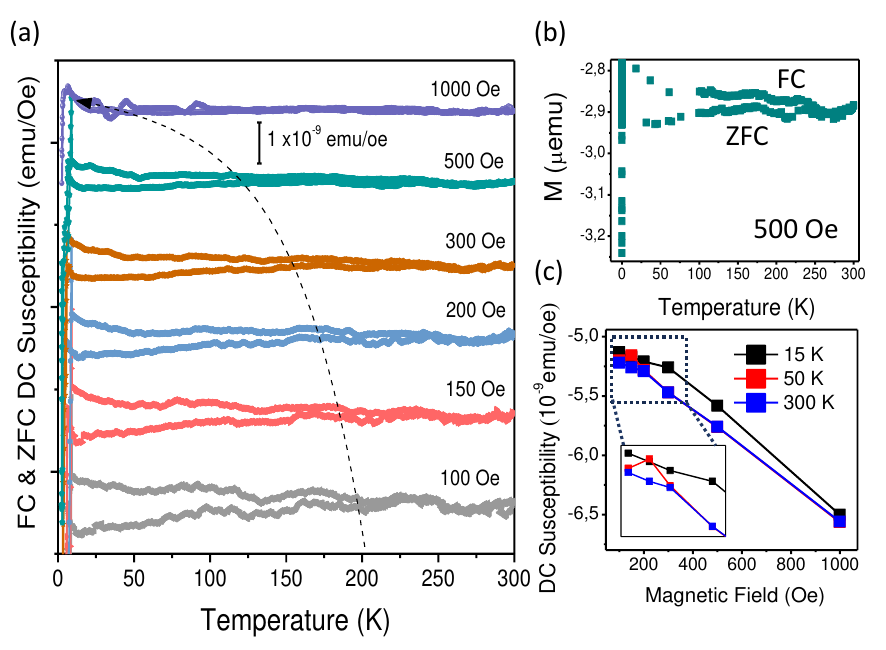}
    \caption{(a) DC susceptibility as a function of temperature under various magnetic fields for a batch of NiBi$_3$ single crystals. Field-cooled (FC) and zero-field-cooled (ZFC) measurements are shown. (b) Magnetization as a function of temperature at 500 Oe. (c) DC susceptibility as a function of magnetic field at 15 K, 50 K, and 300 K for FC measurements. }
    \label{fig:squid_magn}
\end{figure}

Magnetization measurements were performed on a bunch of NiBi$_3$ single crystals using a Superconducting Quantum Interference Device (SQUID) magnetometer MPMS-XL from Quantum Design. Fig. \ref{fig:squid_magn}\textcolor{blue}{a} shows the DC magnetic susceptibility as a function of temperature under various applied magnetic fields. The upper branches correspond to field-cooled (FC) measurements, while the lower branches represent zero-field-cooled (ZFC) data. A noticeable bifurcation between FC and ZFC curves marks the onset of magnetic irreversibility, with the irreversibility temperature for each field indicated by dashed lines. Small oscillations with amplitudes on the order of 1×10$^{-9}$ emu are attributed to minor variations in sample positioning during measurement and do not reflect intrinsic properties of the crystals. Fig. \ref{fig:squid_magn}\textcolor{blue}{b} presents the magnetization as a function of temperature under an applied field of 500 Oe, clearly demonstrating the diamagnetic response of the sample.

\begin{figure}[h]
    \centering
    \includegraphics[width=8.5cm]{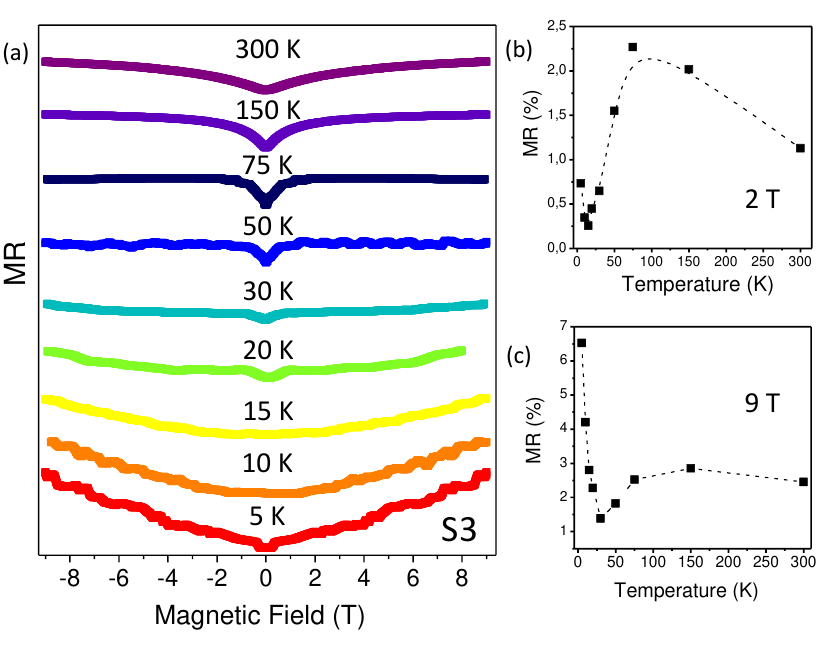}
    \caption{Magnetoresistance (MR) of sample S3. (a) Out-of-scale MR in function of magnetic field for several temperatures. MR values in function of temperature at (b) 2 T and (c) 9 T. }
    \label{fig:MR-S3}
\end{figure}

Magnetoresistance (MR) measurements as a function of magnetic field at various temperatures for sample S3 are shown in Fig. \ref{fig:MR-S3}\textcolor{blue}{a}. From 5 K to 15 K, the MR exhibits a purely quadratic dependence, consistent with ordinary MR mechanisms. At 20 K, a weak linear contribution begins to emerge and becomes increasingly prominent at higher temperatures. As shown in Fig. \ref{fig:MR-S3}\textcolor{blue}{b}, where the MR at 2 T is plotted versus temperature, this linear component reaches a maximum around 75 K and subsequently diminishes. Consequently, as depicted in Fig. \ref{fig:MR-S3}\textcolor{blue}{c}, the overall MR increases with temperature despite a suppression of the quadratic contribution, due to the dominance of the linear term at intermediate temperatures.

\subsection*{Magnetoresistance for Two-Bands Parallel Conduction with Spin Mixing}

Following the seminal work of Fert and Campbell on spin-dependent transport in ferromagnetic alloys \cite{A_Fert_1976}, the total electrical resistivity in the presence of spin-mixing processes can be expressed as
\begin{equation}
    \rho = \frac{\rho_{\uparrow} \rho_{\downarrow} + \rho_{\uparrow\downarrow} (\rho_{\uparrow} + \rho_{\downarrow})}{\rho_{\uparrow} + \rho_{\downarrow} + 4\rho_{\uparrow\downarrow}} \label{eqn:fert_model}
\end{equation}
where $\rho_{\uparrow}$ and $\rho_{\downarrow}$ denote the partial resistivities associated with spin-up and spin-down conduction channels, and $\rho_{\uparrow\downarrow}$ accounts for spin-flip scattering events that couple the two spin channels. 

We define the resistivity spin-channel asymmetry ratio
\begin{equation*}
    \mu = \frac{\rho_{\downarrow}}{\rho_{\uparrow}}.
\end{equation*}

Suppose that Spin mixing resistivity is proportional to the applied magnetic field (H), $\rho_{\uparrow\downarrow}$ = $\alpha$H. Therefore the  magnetoresistance can be calculated as

\begin{align*}
    \Delta \rho(H) & = \rho(H) - \rho(0) = \rho_{\uparrow} \cdot \left[ \frac{ \mu \rho_{\uparrow} + \alpha H (1 + \mu)}{\rho_{\uparrow}(1 + \mu) + 4\alpha H }\right] - \frac{\mu\rho_{\uparrow}}{1 + \mu} 
\end{align*}
After some algebra, one can obtain
\begin{align*}
     \Delta \rho(H) = \rho_{\uparrow} \cdot \frac{ \alpha H (\mu - 1)^2 }{ \rho_{\uparrow}(1 + \mu) + 4 \alpha H}.
\end{align*}

Rewriting in terms of $\rho(0)$, we can simplify to:

\begin{equation}
    \frac{ \Delta \rho(H) }{ \rho(0) } = C \cdot \frac{H}{H_0 + H} \label{eqn:MR_fert_sm},
\end{equation}
where
\begin{equation*}
    C = \frac{(\mu - 1)^2}{4\mu} \qquad\text{and}\qquad H_0 = \frac{(\mu + 1)^2 \rho(0)}{4\mu\alpha}.
\end{equation*}

For low magnetic fields ($H \ll H_0$), Eq. \ref{eqn:MR_fert_sm} can be expanded, leading to
\begin{equation*}
    \Delta \rho(H\ll H_0) \approx C\cdot\rho(0)\cdot\frac{H}{H_0 }\left(\frac{1}{1 + \frac{H}{H_0}}\right)
\end{equation*}
Considering only the linear term, the magnetoresistance becomes
\begin{equation}
    \Delta \rho(H \ll H_0) \approx \alpha \left( \frac{\mu - 1}{\mu + 1} \right)^2 H. \label{eqn:fert_lowfield}
\end{equation}

In the opposite limit of large magnetic field, where $ H \gg H_0$, Eq. \eqref{eqn:MR_fert} simplifies to
\begin{equation*}
    \Delta \rho(H \gg H_0) = C\rho(0)\lim_{H \to \infty} \frac{H}{H_0 + H},
\end{equation*}
which saturates in
\begin{align}
    \Delta\rho(H \gg H_0) & = C\cdot\rho(0) \label{eqn}.
\end{align}

The central hypothesis in this modified Campbell-Fert model is the assumption of a linear relation between the mixing resistivity term and the external magnetic field. Qualitatively, we argue that in the fluctuation regime, spin fluctuation arises from a non-magnetic ground state, on a time scale larger than the carriers scattering. Magnetic correlation between the fluctuations separates the conduction channels into two components, giving rise to a short-range magnetic ordering which spam the AHE and LMR. The external field promotes an asymmetrical enhancement of the spin-flip by forcing the fluctuating moments to align with the field and increasing the mixing resistivity term. 

\begin{figure}[H]
    \centering
    \includegraphics[width=9cm, height=5cm]{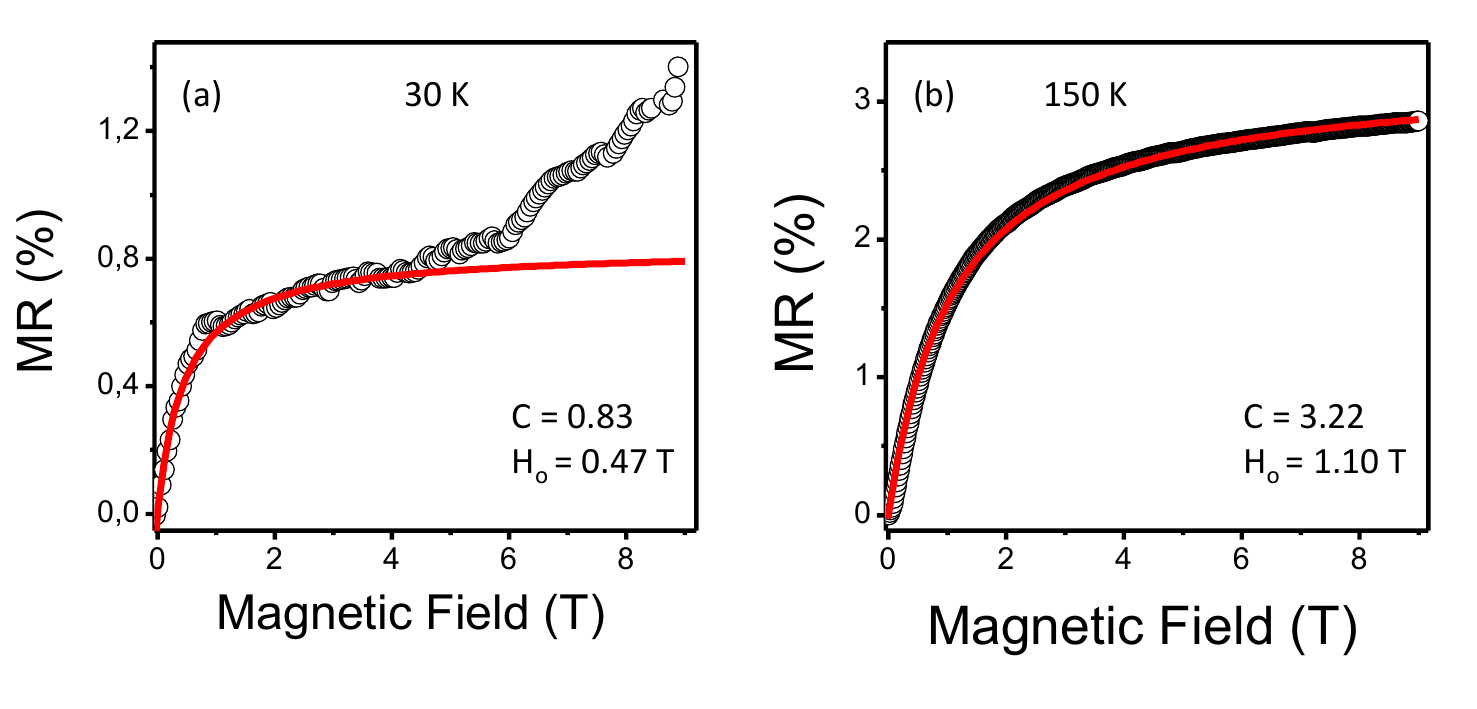}
    \caption{Magnetoresistance as a function of magnetic field for S3 at (a) 30 K and (b) 150 K. The red line corresponds to Eq. \ref{eqn:fert_model}. Model parameters are displayed in the panels.
}
    \label{fig:enter-label}
\end{figure}

30 K has $\mu$ = 5.04 and $\alpha$ = 131.3
150 K has $\mu = 13.93$ and $\alpha = 559.9$

\end{document}